%% file: ms.tex
\documentclass[11pt,final]{article}

\usepackage{fullpage}

\usepackage{microtype}
\usepackage{graphicx}
\usepackage{booktabs} % for professional tables
\usepackage[a4paper,margin=1in]{geometry}

\usepackage{algorithm}
\usepackage{algorithmic}

\usepackage{hyperref}

\usepackage{amsmath}
\usepackage{amssymb}
\usepackage{mathtools}
\usepackage{amsthm}

\usepackage[capitalize,noabbrev]{cleveref}

\theoremstyle{plain}
\newtheorem{theorem}{Theorem}[section]
\newtheorem{proposition}[theorem]{Proposition}
\newtheorem{lemma}[theorem]{Lemma}
\newtheorem{corollary}[theorem]{Corollary}
\newtheorem{example}[theorem]{Example}
\newtheorem{observation}[theorem]{Observation}
\theoremstyle{definition}
\newtheorem{definition}[theorem]{Definition}

\newtheorem{remark}[theorem]{Remark}

\usepackage{enumitem}

\usepackage[numbers,compress]{natbib}

\newenvironment{proofof}[1]{\begin{proof}[\textnormal{\textbf{Proof of \Cref{#1}}}]}{\end{proof}} 

\newenvironment{sketch}[1]{\begin{proof}[\textnormal{\textbf{Proof sketch of \Cref{#1}}}]}{\end{proof}}

\newcommand\ceil[1]{\left\lceil#1\right\rceil}

\DeclareMathOperator*{\argmax}{arg\,max}

\newcommand{\gamename}{\textnormal{MSB}}
\newcommand{\eogame}{\textnormal{EBSB}}
\newcommand{\domeq}{dominant}
\newcommand{\domeqn}{\textnormal{\domeq}}
\newcommand{\onevec}{\textbf{1}}
\newcommand{\zerovec}{\textbf{0}}
\newcommand\abs[1]{\left\lvert#1\right\rvert}

\newcommand{\pog}{PoG}

\usepackage{pgfplots}
\usepackage{dsfont}

\usepackage{pgfplots}
\usepackage{nicefrac}
\usepackage{comment}
\usepackage{tikz}
\usepackage{breakurl}  % For natbib compatibility
\usepackage{xurl}      % Improved line-breaking for URLs

\newcommand{\appnx}[1]{{\ifnum\Includeappendix=1{#1}\else{the appendix}\fi}}

\newcount\Includeappendix 
\usepackage{subcaption}
\usepackage[textsize=tiny]{todonotes}

\title{Collaborating with GenAI: Incentives and Replacements}
\author{
Boaz Taitler%
\thanks{%
    {Technion---Israel Institute of Technology (\url{boaztaitler@campus.technion.ac.il})}}
\and Omer Ben{-}Porat%
\thanks{%
    {Technion---Israel Institute of Technology (\url{omerbp@technion.ac.il})}}
}

\begin{document}

\maketitle

\begin{abstract}
The rise of Generative AI (GenAI) is reshaping how workers contribute to shared projects. While workers can use GenAI to boost productivity or reduce effort, managers may use it to replace some workers entirely. We present a theoretical framework to analyze how GenAI affects collaboration in such settings. In our model, the manager selects a team to work on a shared task, with GenAI substituting for unselected workers. Each worker selects how much effort to exert, and incurs a cost that increases with the level of effort. We show that GenAI can lead workers to exert no effort, even if GenAI is almost ineffective. We further show that the manager's optimization problem is NP-complete, and provide an efficient algorithm for the special class of (almost-) linear instances. Our analysis shows that even workers with low individual value may play a critical role in sustaining overall output, and excluding such workers can trigger a cascade. Finally, we conduct extensive simulations to illustrate our theoretical findings.
\end{abstract}

\input{Sections/Introduction.tex}

\input{Sections/model.tex}

\input{Sections/GenAINegativeEffect}
\input{Sections/OptimalCoalition}
\input{Sections/Stability}
\input{Sections/simulation}
\input{Sections/discussion.tex}

\section*{Acknowledgements}
This research was supported by the Israel Science Foundation (ISF; Grant No. 3079/24).

\bibliography{ms}
\bibliographystyle{abbrvnat}

%%%%%%%%%%%%%%%%%%%%%%%%%%%%%%%%%%%%%%%%%%%%%%%%%%%%%%%%%%%%%%%%%%%%%%%%%%%%%%%
%%%%%%%%%%%%%%%%%%%%%%%%%%%%%%%%%%%%%%%%%%%%%%%%%%%%%%%%%%%%%%%%%%%%%%%%%%%%%%%
% APPENDIX
%%%%%%%%%%%%%%%%%%%%%%%%%%%%%%%%%%%%%%%%%%%%%%%%%%%%%%%%%%%%%%%%%%%%%%%%%%%%%%%
%%%%%%%%%%%%%%%%%%%%%%%%%%%%%%%%%%%%%%%%%%%%%%%%%%%%%%%%%%%%%%%%%%%%%%%%%%%%%%%
\newpage

{\ifnum\Includeappendix=1{
\appendix
\onecolumn
\input{appendix}
}{\fi}

%%%%%%%%%%%%%%%%%%%%%%%%%%%%%%%%%%%%%%%%%%%%%%%%%%%%%%%%%%%%%%%%%%%%%%%%%%%%%%%
%%%%%%%%%%%%%%%%%%%%%%%%%%%%%%%%%%%%%%%%%%%%%%%%%%%%%%%%%%%%%%%%%%%%%%%%%%%%%%%

\end{document}

%% file: Sections/Introduction.tex
\section{Introduction}

Generative AI has transformed how individuals approach their work, allowing workers to create and refine content with significantly less time and effort. The growing availability of powerful generative AI tools at minimal cost has made them widely accessible and increasingly embedded in everyday tasks. As a result, workers often rely on these tools either to assist them or to complete parts of the work on their behalf. This development raises an important question: how much effort are workers still contributing, and how much of the work is effectively being handled by AI?

When applied to a new task, generative AI can serve two distinct roles. First, it can augment a worker's effort, enabling them to contribute more effectively than they would unaided. Second, it can substitute for parts of the worker's labor, allowing them to engage less or even not at all with certain aspects of the task. This distinction has implications for concerns about misuse and over-reliance. In educational settings, for example, students may use generative AI to complete assignments without engaging with the material itself~\cite{reiter2025student, cotton2024chatting}. In hiring, excessive reliance on AI-driven tools has been linked to the amplification of existing biases~\cite{chen2023ethics}.

Consider a scenario where a manager must assemble a team to work on a shared project, with the outcome to be divided among the workers and the manager. If AI can surpass a worker's output, or if the worker chooses to rely primarily on AI without investing meaningful effort, the manager may prefer to replace that worker with AI, retaining a larger share of the benefit. Growing attention highlights how AI is beginning to replace human workers, with early cases of layoffs linked to generative AI~\cite{washingtonpost_ibm_ai_2023, wsj_ibm_ai_2025, bbc_ibm_ai_2023}. These developments raise growing concerns about how the availability of generative AI is reshaping incentives around effort, inclusion, and replacement.

Although there is a growing body of work on the capabilities and performance of generative tools, much less is known about how access to these tools affects workers' incentives to exert effort, or how replacing a worker with AI impacts the overall outcome and the behavior of remaining workers. This prompts two critical questions: Will individuals continue to invest effort when they can offload tasks to AI? And should those coordinating group efforts replace some workers with generative AI, and if so, whom?

\paragraph{Our contribution} Our contribution is twofold. The first is conceptual: we present a stylized model that captures team interplay with generative AI. There are two types of participants in our game. The first type represents workers working on a project that generates revenue, or contributors collaborating on a shared product. We refer to this type more generally as players. Each player chooses how much effort to invest, incurring a corresponding cost, and whether to use generative AI to assist with their work. The players collaborate on a shared task that produces a collective benefit, which is then divided among them. Importantly, the productivity of each player increases the total benefit, thereby improving the outcome for all participants.

The other participant type is the manager, who selects a subset of players to work on the task. The manager receives a portion of the shared benefit and may also utilize generative AI to perform the tasks of players who are not selected. Motivated by the widespread availability of powerful commercial tools such as ChatGPT and Gemini offered under free or low-cost plans, we assume that generative AI is freely accessible to both the manager and the players.

Our second contribution is technical. We begin by comparing player behavior under two settings: one in which players cannot use generative AI, and another in which they are required to use it. This comparison shows that in equilibrium, generative AI can incentivize players to exert no effort, even when GenAI's effectiveness is negligible.

We then turn to the computational challenge of identifying the optimal coalition from the manager's perspective. We show that this problem is NP-complete and provide an efficient algorithm for the special case where the collective benefit is nearly linear in the players' contributions. We next analyze the stability of the optimal coalition and show that it is brittle: it may include players whose marginal contribution is nearly zero, yet whose removal is detrimental. Finally, we conduct simulations to analyze the properties of optimal coalitions. We demonstrate that coalitions tend to be either large or small, but rarely medium.

\subsection{Related Work}
Our work is inspired by the emerging body of research at the intersection of foundation models and game theory~\cite{raghavan2024competition, laufer2024fine, conitzerposition, dean2024accounting}. This literature examines the impact of generative AI on user welfare and utility in the context of competition~\cite{taitler2025braess, keinan2025strategic}, content diversity~\cite{raghavan2024competition}, and applications inspired by social choice theory~\cite{boehmer2025generative} and mechanism design~\cite{conitzerposition, sun2024mechanism}.

From a conceptual perspective, our work relates to the literature on coalition formation~\cite{aumann1974cooperative}, which encompasses both strategic considerations for forming stable coalitions~\cite{bilo2022hedonic, taguelmimt2023optimal} and algorithmic approaches to learning optimal coalitions~\cite{cohen2025online, singla2015learning}. Related themes also appear in the literature on credit assignment and resource allocation among agents~\cite{cookson2025constrained, barman2025fair, bredereck2022improving}.

Our work is also related to the literature on public goods games~\cite{ledyard1994public, kollock1998social}, which model scenarios where multiple players contribute to a shared cause while incurring a personal cost. This line of work includes studies on strategic behavior and free-riding~\cite{cheng2025networked}, computational aspects of equilibrium computation~\cite{darvariu2021solving}, and incentive mechanisms to promote altruistic behavior~\cite{altruism_public_good}. Delegation games~\cite{oesterheld2022safe, coop_control}, where a human delegates decision-making authority to an AI agent, is also close in spirit. Also related is the line of work on deferring tasks to AI~\cite{goktas2025strategic, noti2022learning, hemmer2023learning, greenwood2025designing, peng2025no}, which is often linked to concerns about AI alignment and robustness to human preferences~\cite{procaccia2025clonerobust}.

From a broader perspective, recent research uses lab experiments to study the effects of AI on working teams~\cite{peng2023impact, becker2025measuring}. The former highlights AI's potential to increase productivity, while the latter finds that although AI is perceived as a productivity-enhancing tool, it can actually reduce productivity. Our paper focuses on a theoretical perspective that is naturally limited, but provides insights on the dynamics of AI use in team settings.

%% file: Sections/model.tex
\section{Model} \label{sec:model}

We study a game involving a manager that we call Principal and $N$ players, which we refer to as the Managed Shared Benefit game ($\gamename$). Formally, the game is defined as $G = (\mathcal{N}, \prod_{i = 1}^N A_i, (\theta_i)_{i = 1}^N, (s_i)_{i = 1}^N, (c_i)_{i = 1}^N, F)$, where the set of players is $\mathcal{N} = \{1, \ldots, N\}$. The action space of each player $i$ is $A_i = [0,1] \times \{ 0, 1 \}$ for which each player chooses an action $(e_i, g_i) \in A_i$, where $e_i$ represents the player's effort level and $g_i$ indicates whether the player chooses to use generative AI.  Principal chooses a coalition $\mathcal{C}$, which is a subset of players $\mathcal{C} \subseteq \mathcal{N}$. A strategy profile is given by $(\textbf{e}, \textbf{g}, \mathcal{C})$, where $\textbf{e} = (e_1, \ldots, e_N)$ and $\textbf{g} = (g_1, \ldots, g_N)$ represent the effort levels and GenAI usages of all players, respectively.

The utility of each player $i \in \mathcal N$ is composed of two components: a benefit derived from a shared resource that depends on all players' actions, and an individual cost that depends on the player's own effort. We now describe each of those components.

\paragraph{Cost functions} Each player incurs an individual cost $c_i : [0,1] \rightarrow \mathbb{R}$, which depends only on their effort level $e_i$. We assume, motivated by the prevalence of GenAI tools, that players face no direct cost for using GenAI. The cost function is assumed to be increasing in effort.

\paragraph{Shared benefit and contribution functions}
Each player's action yields a task-specific output determined by a \emph{contribution function} $s_i : A_i \rightarrow \mathbb{R}$, which maps effort and GenAI usage to a scalar contribution value. The function $s_i(e_i, g_i)$ reflects how effective player $i$'s effort is when assisted by GenAI. This mapping further captures the effectiveness of GenAI toward the contribution. For example, in tasks where GenAI is highly relevant, using GenAI can amplify productivity. In contrast, when GenAI lacks domain-specific data, its use may have little or no effect on the contribution. We assume that the contribution function is nondecreasing in both effort and GenAI usage.

The contributions of all players collectively generate a nonnegative \emph{shared benefit}, modeled by a function $F : \mathbb{R}^N \rightarrow \mathbb{R}_{\geq 0}$. The shared benefit $F(s_1, \ldots, s_N)$ represents the overall output of the project, such as the success of a shared endeavor or the quality of a final product, which is distributed among the participants.

In the game, only players in the coalition chosen by Principal contribute actively to the shared benefit. Players who are not selected are replaced by GenAI. We define $\tilde{s}_i(e_i, g_i, \mathcal{C}) = \mathds{1}_{\{i \in \mathcal{C}\}} s_i(e_i, g_i) + \mathds{1}_{\{i \notin \mathcal{C}\}} s_i(0, 1)$ as the coalition-dependent contribution of player $i$: the players' contribution under $e_i, g_i$ if that player is in the coalition, or the contribution of GenAI otherwise. The coalition-dependent shared benefit is then given by
\[
f(\textbf{e}, \textbf{g}, \mathcal{C}) = F\left( \tilde{s}_1(e_1, g_1, \mathcal{C}), \ldots, \tilde{s}_N(e_N, g_N, \mathcal{C}) \right).
\]

\paragraph{Utilities}
Each player selected by Principal receives a fixed portion $\theta_i \in [0, 1]$ of the shared benefit. These portions are predetermined and remain fixed throughout the game, with no possibility of utility transfer between players.\footnote{This approach allows us to provide insight to how players respond in the game to their assigned portions, which is a step toward understanding how credit should be assigned and its effects on the outcomes} Principal's share of the benefit is the remaining portion, given by $1 - \sum_{i \in \mathcal{C}} \theta_i$.

Overall, the utility of each player $i \in \mathcal{N}$ is given by
\[
U_i(\textbf{e}, \textbf{g}, \mathcal{C}) = \mathds{1}_{\{ i \in \mathcal{C} \}} \cdot \theta_i f(\textbf{e}, \textbf{g}, \mathcal{C}) - c_i(e_i),
\]
and the utility of Principal is defined as
\[
W(\textbf{e}, \textbf{g}, \mathcal{C}) = \left(1 - \sum_{i \in \mathcal{C}} \theta_i \right) f(\textbf{e}, \textbf{g}, \mathcal{C}).
\]

\subsection{Structure of the shared benefit}
We focus on a natural and expressive subclass of shared benefit functions: multilinear polynomial functions.

\begin{definition}[Multilinear polynomial function]
A function $h : \mathbb{R}^N \rightarrow \mathbb{R}$ is multilinear polynomial if for every subset $\mathcal{X} \subseteq [N]$, there exists a coefficient $\gamma_{\mathcal{X}} \in \mathbb{R}_{\geq 0}$ such that
\[
h(x_1, \ldots, x_N) = \sum_{\mathcal{X} \subseteq [N]} \gamma_{\mathcal{X}} \prod_{i \in \mathcal{X}} x_i.
\]
\end{definition}
Multilinear functions capture a wide range of interaction patterns among players. The additive structure reflects independent contributions. The multiplicative terms represent complementarities, where value emerges only when several players contribute together. For example, in a software project, both frontend and backend components may be required for the system to work. If either one is missing, the product may fail completely, which reflects a strong interaction between efforts.

\subsection{Example}
In this subsection, we introduce a concrete instance of a $\gamename$ game, which serves as a running example throughout the paper.

\begin{example} \label{example}
\normalfont
Consider a game with two players and Principal, where each player receives a fixed portion of the shared benefit: $\theta_1 = \theta_2 = 0.3$. The shared benefit function is given by $F(s_1, s_2) = 8 s_1 s_2$, and the contribution functions are defined as $s_1(e_1, g_1) = \sqrt{e_1}$ and $s_2(e_2, g_2) = \sqrt{e_2 + 0.2 g_2}$. The players incur effort costs specified by $c_1(e_1) = \ln(1 + e_1)$ and $c_2(e_2) = 3 \ln(1 + e_2)$. The utilities of the players are given by
{\small
\begin{align*}
    U_1(\textbf{e}, \textbf{g}, \mathcal{C}) = \begin{cases}
        2.4 \sqrt{e_1} \sqrt{e_2 + 0.2 g_2} - \ln(1+e_1) & \mbox{$\mathcal{C} = \mathcal{N}$} \\
        2.4\sqrt{0.2} \sqrt{e_1} - \ln(1+e_1) & \mbox{$\mathcal{C} = \{ 1 \}$} \\
        0 - \ln(1+e_1) & \mbox{$1 \notin \mathcal{C}$} 
    \end{cases} ,
\end{align*}
\begin{align*}
    U_2(\textbf{e}, \textbf{g}, \mathcal{C}) = \begin{cases}
        2.4 \sqrt{e_1} \sqrt{e_2 + 0.2 g_2} - 3\ln(1+e_2) & \mbox{$\mathcal{C} = \mathcal{N}$} \\
        0 - 3\ln(1+e_2) & \mbox{$1 \notin \mathcal{C}$}
    \end{cases} ,
\end{align*}
}%
and the utility of Principal is
\begin{align*}
    W(\textbf{e}, \textbf{g}, \mathcal{C}) = \begin{cases}
        3.2 \sqrt{e_1} \sqrt{e_2 + 0.2 g_2} & \mbox{$\mathcal{C} = \mathcal{N}$} \\
        5.6 \sqrt{0.2} \sqrt{e_1}  & \mbox{$\mathcal{C} = \{ 1 \}$} \\
        0 & \mbox{$1 \notin \mathcal{C}$}
    \end{cases} .
\end{align*}

In this example, Principal selects one of the coalitions $\{1, 2\}$, $\{1\}$, $\{2\}$, or the empty set $\emptyset$, while the players choose their effort and GenAI usage levels $(e_1, e_2, g_1, g_2)$. Notably, GenAI does not improve player 1's contribution, making player 1 indifferent between using GenAI or not. Moreover, the simultaneous move version of this game, where Principal and the players choose their actions at the same time, has no equilibrium. Specifically, for any effort level of Player 2 when using GenAI, player 1's best response is to exert full effort $e_1 = 1$ if included in the coalition. On the other hand, player 2 has a dominant strategy to use GenAI and to invest no effort.

\end{example}

\subsection{Further Notations}
We now introduce the notations used throughout this work. Given a vector $\mathbf{V} \in \mathbb{R}^N$, we denote by $\mathbf{V}_{-i}$ the vector in $\mathbb{R}^{N-1}$ that contains all entries of $\mathbf{V}$ except for the $i$-th entry $V_i$. Finally, for two vectors $\mathbf{A}, \mathbf{B} \in \mathbb{R}^N$, the notation $\mathbf{A} > \mathbf{B}$ and $\mathbf{A} \geq \mathbf{B}$ indicate that for every element $i \in \{1, \ldots, N\}$, it holds that $A_i > B_i$ and $A_i \geq B_i$, respectively. We further denote by $\textbf{0} = (0, \ldots, 0)$ and $\textbf{1} = (1, \ldots, 1)$ the vectors in $\mathbb{R}^N$ whose entries are all zero and one, respectively.

%% file: Sections/GenAINegativeEffect.tex
\section{GenAI's Negative Effect on Effort} \label{sec: free ride}
In this section, we analyze the effect of allowing players to use GenAI, showing that it can reduce the effort they choose to invest. We fix a coalition of players selected by Principal and compare their behavior in two settings: one in which players are restricted from using GenAI, and another in which GenAI is freely accessible. This comparison reflects the transition from a time when GenAI was not available to a more modern setting where it is commonly used as part of the work process.

To study this effect formally, we narrow our attention to the class of  sub-games where both the coalition and GenAI usage profile are fixed, where players choose only their effort levels. We term this class \emph{Effort-based Shared-Benefit games} ($\eogame$).

Formally, given an $\gamename$ game, a GenAI usage profile $\textbf{g}$ and coalition $\mathcal{C}$, the corresponding $\eogame$ game $\Tilde{G}=\Tilde{G}(G, \textbf{g}, \mathcal{C})$ is a game where each player chooses an effort $e_i \in [0, 1]$ and the utility is given by $U_i(\textbf{e}, \textbf{g}, \mathcal{C})$. For brevity, when $\textbf{g}$ and $\mathcal{C}$ are clear from the context, we let $\tilde U_i(\textbf{e}) = U_i(\textbf{e}, \textbf{g}, \mathcal{C})$.

Throughout this section, we focus on two specific instances of the $\eogame$ games: one where $\textbf{g} = \zerovec$, meaning no player uses GenAI, and one where $\textbf{g} = \onevec$, which represents the case where players use GenAI.

\subsection{Dominant Equilibrium in $\eogame$ games} \label{subsec: solution concept}
In this subsection, we highlight a special kind of Pure Nash Equilibrium (hereafter PNE) in $\eogame$ games. To do so, we first describe the concept of supermodular games~\cite{topkis1998supermodularity}. For convenience, we shall present a sufficient condition for a game to be supermodular in terms of $\eogame$ games, and defer the formal definition of supermodular games to \appnx{Appendix~\ref{appn: free ride}}.
\begin{definition}
We say an $\eogame$ game is supermodular if for every player $i$ and every profiles $\textbf{e}$ and $\textbf{e}'$ such that $\textbf{e} \geq \textbf{e}'$, it holds that 
\begin{align*}% \label{eq:sup body}
    \tilde U_i(e_i, \textbf{e}_{-i}) -  \tilde U_i(e'_i, \textbf{e}_{-i}) \geq \tilde U_i(e_i, \textbf{e}'_{-i}) - \tilde U_i(e'_i, \textbf{e}'_{-i}).
\end{align*}
\end{definition}
% A supermodular game is a special class of lattice games, defined as follows:
% \begin{definition}
% Let $(\mathcal{N}, S \{ f_i : i \in \mathcal{N}\})$ be a game where $\mathcal{N}$ is the set of players, $S = \times_{i \in \mathbb{N}}S_i$ is the space of feasible joint strategies and $f_i : S \rightarrow \mathbb{R}$ is the payoff of player $i$. The game is supermodular if:
% \begin{enumerate}
%     \item $S$ is a sublattice of $\times_{i \in \mathbb{N}} \mathbb{R}^{m_i}$.
%     \item $f_i(y_i, x_{-i})$ is supermodular in $y_i$ on $S_i$ for each $x_{-i}$ on $S_{-i}$ and every player $i$.
%     \item $f_i(y_i, x_{-i})$ has increasing differences in $(y_i, x_{-i})$ on $S_i \times S_{-i}$.
% \end{enumerate}
% \end{definition}

Using the multilinear polynomial structure of the shared benefit function, we show that:
\begin{proposition} \label{supermodular game}
Any $\eogame$ game is a supermodular game.
\end{proposition}
Establishing that every $\eogame$ game is supermodular is particularly useful since it guarantees the existence of at least one pure Nash equilibrium~\cite{vives1990nash}. Moreover, the set of pure equilibrium profiles is ordered: any two equilibria can be compared element-wise.
\begin{corollary}
Every $\eogame$ game $\Tilde{G}$ admits at least one PNE. Furthermore, if $\textbf{e}, \textbf{e}'$ are two PNEs of $\Tilde{G}$, then either $\textbf{e} \geq \textbf{e}'$ or $\textbf{e} \leq \textbf{e}'$.
\end{corollary}
Furthermore, due to the monotonicity of the shared benefit function, we show that the order of equilibria applies to player utilities as well.
%In our setting, the supermodularity arises from the properties of the shared benefit function. Since this function is non-decreasing in each argument, a player's incentive to exert effort increases when others do the same. This allows us to further rank the utilities across different equilibrium profiles.
\begin{lemma} \label{obs: increasing equilibria}
Let $\textbf{e}$ and $\textbf{e}'$ be pure Nash equilibrium profiles such that $\textbf{e} \geq \textbf{e}'$. Then, for every player $i \in \mathcal{N}$, it holds that $u_i(\textbf{e}; \textbf{g}, \mathcal{C}) \geq u_i(\textbf{e}'; \textbf{g}, \mathcal{C})$.
\end{lemma}

\Cref{supermodular game} and \Cref{obs: increasing equilibria} imply that there exists a unique pure equilibrium that is socially optimal and Pareto dominates all other pure equilibria. We refer to this profile as the \emph{dominant equilibrium}. We denote it by $\sigma^\star(\textbf{g}, \mathcal{C})$, representing the $\domeqn$ equilibrium in the $\eogame$ game induced by GenAI profile $\textbf{g}$ and coalition $\mathcal{C}$. Given an $\gamename$ game, we use the dominant equilibrium as our solution concept and as a benchmark for comparing different $\eogame$ games. We also remark that finding the dominant equilibrium could be done by employing a best-response dynamics with a carefully chosen initial profile~\cite{topkis1998supermodularity}.

\subsection{Low effort inducing GenAI} \label{subsec: low effort inducing}
We now examine whether the option to use GenAI can lead players to reduce their effort. To address this question, we compare $\eogame$ games with the same coalition under two scenarios: one where players are restricted from using GenAI, and one where GenAI is available.

Recall that by definition, $\sigma^\star(0, \mathcal{C})$ is the effort profile in equilibrium of the players when they cannot use GenAI. Furthermore, the contribution of each player is non-decreasing in that player's usage of GenAI. Therefore, allowing the players to deviate and use GenAI can only make everyone better off. Surprisingly, GenAI can lead to the opposite effect, inducing players to misuse it and contribute less instead.
\begin{proposition} \label{obs: example compare}
There are $\gamename$ games and coalition $\mathcal{C} \subseteq N$ such that $s_i(\sigma_i^\star(0, \mathcal{C}), 0) \geq s_i(\sigma_1^\star(1, \mathcal{C}), 1)$ for every $i \in \mathcal{C}$ and there is $i \in \mathcal{C}$ where $s_i(\sigma_i^\star(0, \mathcal{C}), 0) > s_i(\sigma_1^\star(1, \mathcal{C}), 1)$.
\end{proposition}
\Cref{obs: example compare} implies that allowing players to use GenAI can make them selfish and be detrimental to the shared benefit and to Principal's utility. Furthermore, it also implies that GenAI prompts players to invest less effort compared to the situation where they could not use GenAI.
\begin{sketch}{obs: example compare}
We use Example~\ref{example} to show \Cref{obs: example compare}. We compare the effort profiles under the coalition $\mathcal{C} = \mathcal{N}$ when $\textbf{g} = \zerovec$ and $\onevec$ respectively. Whenever the contribution of the second player is $s_2(e_2, g_2) \geq \sqrt{0.2}$, the first player invests $e_1 = 1$, leading to the $\domeqn$ equilibrium $\sigma^\star(\zerovec, \mathcal{C}) = (1, 0.25)$. Next, notice that allowing the players to use GenAI ensures that $s_2(e_2, 1) \geq \sqrt{0.2}$ and therefore the first player does not deviate. In contrast, GenAI allows the second player to substantially cut costs and therefore has the incentive to deviate, leading to the $\domeqn$ equilibrium $\sigma^\star(\onevec, \mathcal{C}) = (1, 0)$.
\end{sketch}

To understand why or when this can occur, it is helpful to examine the marginal utility each player receives from increasing their effort. This marginal utility consists of two components: a positive term that reflects the gain in shared benefit due to higher contribution, and a negative term representing the additional cost of effort. A key insight is that players are more inclined to exert effort when their marginal contribution is high. Thus, if GenAI lowers a player's marginal contribution, their incentive to invest effort may diminish. Following this intuition, one might expect that when GenAI has only a minimal effect on the contribution, the change in effort would be small. Surprisingly, this is not the case: even when GenAI alters the contribution by an arbitrarily small amount, it can cause a complete collapse of effort in equilibrium.
\begin{proposition} \label{GenAI induced no effort}
For any $\varepsilon > 0$, there exists an $\gamename$ game, coalition $\mathcal{C}$ and a player $i \in \mathcal{C}$ such that $\left| s_i(e_i, 1) - s_i(e_i, 0)\right|  \leq \varepsilon$ and it holds that $\sigma_i^\star(\zerovec, \mathcal{C}) = 1$ and $\sigma_i^\star(\onevec, \mathcal{C}) = 0$.
\end{proposition}

\subsection{Price of GenAI} \label{subsec: price of GenAI}

We now define the \emph{\textbf{p}rice \textbf{o}f \textbf{G}enerativity} (PoG) as the inefficiency introduced by allowing players to use GenAI. The PoG quantifies the ratio between the shared benefit achieved when players are restricted from using GenAI and the shared benefit when they are allowed to use it. Formally, consider a fixed coalition $\mathcal{C}$ and the corresponding $\eogame$ games induced by GenAI usage profiles $\zerovec$ and $\onevec$. The PoG is defined by
\[
\pog = \frac{f(\sigma^\star(\zerovec, \mathcal{C}), \zerovec, \mathcal{C})}{f(\sigma^\star(\onevec, \mathcal{C}), \onevec, \mathcal{C})}.
\]

Recall that \Cref{GenAI induced no effort} showed that players might choose to exert no effort even when GenAI provides only a negligible improvement to their contribution. This observation is critical because it implies that players may settle for minimal contributions in equilibrium. The lower these contributions are, the greater the potential harm to the shared benefit.

\begin{proposition}\label{pog result}
For every $\varepsilon > 0$, there exists an instance where $\left| s_i(e_i, 1) - s_i(e_i, 0)\right| \leq \varepsilon$ for every player $i \in \mathcal{C}$, and it holds that $\pog < \frac{1}{\sqrt{\varepsilon}}$.
\end{proposition}

This result demonstrates that even when the benefit of using GenAI is arbitrarily small for each player, the collective outcome can be significantly worse. The inefficiency introduced by GenAI does not depend solely on how much it improves individual contributions, but also on how it alters players' incentives to invest effort. As a result, even a slight reliance on GenAI can lead to substantial degradation in the shared benefit.

%% file: Sections/OptimalCoalition.tex
\section{Finding The Optimal Coalition} \label{sec: replace}
In this section, we focus on the question of how Principal should choose a coalition, given that Principal acts first. This setting represents a broad class of interactions. For example, in a hiring process, a manager first decides whom to hire, and only afterward do the workers invest their effort. We model this as a Stackelberg game in which Principal acts first by choosing and committing to a coalition, and then the players decide how much effort to invest and whether to use GenAI. Recall that using GenAI is a weakly dominant strategy for the players, as it increases their contribution without incurring any additional cost. Therefore,
\begin{remark}
    Throughout the rest of this paper we assume that $\textbf{g} = \onevec$ and, with slight abuse of notation, omit the GenAI usage profile.
\end{remark}
Principal's objective is to select the coalition that maximizes his own utility, accounting for the players' subsequent behavior.
Fixing a coalition $\mathcal{C}$ induces a corresponding $\eogame$ game. We assume that players respond by playing the $\domeqn$ equilibrium $\sigma^\star(\mathcal{C})$ of that game. Given this, Principal faces the following optimization problem:
\begin{align} \label{find coalition}
&\max_{\mathcal{C} \in \mathcal{N}} W(\sigma^\star(\mathcal{C}), \mathcal{C}). \tag{P1}
\end{align}

\subsection{On the complexity of finding the coalition} \label{subsec: two stage hardness}
As a two stage game, Problem~\eqref{find coalition} can be solved by first computing the $\domeqn$ equilibrium in the $\eogame$ game induced by each possible coalition, and then optimizing over the choice of coalition. The $\domeqn$ equilibrium profile can be found by leveraging the supermodular structure of the $\eogame$ game, which guarantees convergence of best-response dynamics. 

In contrast, optimizing over the coalitions is more subtle. The tradeoff between aiming to maximize the portion of the shared benefit and increasing the overall shared benefit lies at the heart of the difficulty of the problem. On the one hand, Principal wants to maximize the shared benefit. For that, having as many players as possible in the coalition is beneficial as they can contribute more than simply using GenAI. On the other hand, each player in the coalition reduces Principal's portion of the shared benefit. Therefore, Principal decides whether to have a player in the coalition or not based on that player's marginal effect on the shared benefit compared to the portion of the shared benefit that this player takes. The next~\Cref{find coalition np-hard} shows computationally hardness by reducing Problem~\eqref{find coalition} from the clique problem.
\begin{theorem} \label{find coalition np-hard}
Problem~\eqref{find coalition} is NP-complete. 
\end{theorem}

\subsection{Optimal solution for special cases} \label{subsec: alg}
Even though the problem is computationally challenging, we can efficiently compute a solution for special cases. To that end, we introduce another formulation of Problem~\eqref{find coalition} under the following mild assumption. We focus on instances where each player's portion of the shared benefit is a multiple of $\varepsilon$, for some $\varepsilon > 0$. Formally, there exists a small number $\varepsilon$, $\varepsilon>0$ such that for every player $i\in \mathcal N$, there exists a positive integer $k_i, k_i \in \mathbb N$ such that $\theta_i = \varepsilon \cdot k_i$. Therefore, each coalition $\mathcal{C}$ is associated with a natural number $k \in \mathbb{N}$ such that the term $\sum_{i \in \mathcal{C}} \theta_i = k\varepsilon$ represents the portion that Principal forgoes when choosing coalition $\mathcal{C}$. This allows us to formulate Problem~\eqref{find coalition} as a bilevel optimization problem:
% \begin{align}
% &\max_{k \in \mathbb{N}} (1-k\varepsilon) f(\sigma^\star(\mathcal{C}_k), \mathcal{C}_k) \label{outer_eq}\\
% & \text{such that } \quad \mathcal{C}_k \in \argmax_{\mathcal{C} \subseteq \mathcal{N}} f(\sigma^\star(\mathcal{C}), \mathcal{C})  \label{inner_eq}\\
% & \qquad \quad  \quad \sum_{i \in \mathcal{C}_k} \theta_i = k\varepsilon \nonumber
% \end{align}
\begin{align}\label{find budget k}
&\max_{k \in \mathbb{N}} (1-k\varepsilon) f(\sigma^\star(\mathcal{C}_k), \mathcal{C}_k) \tag{P2}\\
& \text{where $\mathcal C_k$ is a solution to:} \nonumber\\
&\qquad \quad \argmax_{\mathcal{C} \subseteq \mathcal{N}} f(\sigma^\star(\mathcal{C}), \mathcal{C}) \label{eq:p2 inner} \tag{P2-inner}\\
& \qquad \quad\text{s.t.} \sum_{i \in \mathcal{C}} \theta_i \leq k\varepsilon \nonumber
\end{align}
\begin{observation}
    Problems~\eqref{find coalition} and \eqref{find budget k} are equivalent.
\end{observation}
Next, we focus on \emph{linear instances}. Specifically, assume that there exist $\gamma_1,\dots,\gamma_N$ such that $F(s_1, \ldots, s_N)=\sum_{i=1}^N \gamma_i s_i$. In such a case, the best response of every player becomes an optimization problem that is independent of the other players' actions. namely, $U_i(\textbf e, \mathcal C) \propto \gamma_i s_i(e_i)-c_i(e_i) +Z $ for some constant $Z$ that does not depend on $e_i$. We shall assume that players can compute best response in $O(1)$. Denote the maximizer of player $i$ by $s_i^\star$. As a result, we can replace the objective function in the inner problem in \eqref{eq:p2 inner} with
\[
f(\sigma^\star(\mathcal{C}), \mathcal{C}) = \sum_{i=1}^N \gamma_i s_i(0) + \sum_{i \in \mathcal C}\gamma_i (s_i^{\star} -s_i(0)).
\]
Crucially, the inner problem reduces to the knapsack problem~\cite{martello1990knapsack}, where we are given a budget $k\varepsilon$ and $N$ items such that each item $i$ has a weight $\theta_i$ and value $\gamma_i (s_i^{\star} -s_i(0))$. This inner problem could be solved in pseudo-polynomial runtime. For the outer problem, iterating over every $k$, $0\leq k \leq \ceil{\nicefrac{1}{\varepsilon}}$ yields an optimal solution.
\begin{theorem} \label{alg for finding coalition}
%Let $\varepsilon > 0$. If $F(s_1, \ldots s_N)$ is a polynomial of degree one and $\theta_i \in \{0, \varepsilon, \ldots 1 \}$ then there exists an algorithm whose output $\mathcal{C}^\star$ is optimal and its runtime is $O(\frac{N}{\varepsilon^2})$.
Assume $F(s_1, \ldots s_N)$ is linear, and further assume that for every $i \in \mathcal N  $, $\theta_i = \varepsilon \cdot k_i$ for some $k_i \in \mathbb N$ and $\varepsilon>0$.  There exists an algorithm that runs in $O(\frac{N}{\varepsilon^2})$ time and returns the optimal solution to Problem~\eqref{find coalition}.
\end{theorem}
We end this section with describing how our approach extends beyond linear shared benefit, for instances that we term \emph{almost linear}. Assume that we can decompose $F$ into a sum such that $F=F_{\mathcal {N'}} + F_{\mathcal N-\mathcal{N'}}$, such that $F_{\mathcal{N'}}$ consists of elements that belong to players in $\mathcal{N'}$ only, and similarly for $ F_{\mathcal N-\mathcal{N'}}$. Furthermore, assume that $\mathcal{N'} << \mathcal N$ and that $F_{\mathcal N-\mathcal{N'}}$ is linear. In such a case, we can exhaustively search over $2^{\abs{\mathcal{N'}}}$ partial coalitions of $\mathcal{N'}$, and for each solve the inner problem~\eqref{eq:p2 inner}, fixing $F_{\mathcal {N'}}$ in $f(\cdot)$ and taking into account their portions. This enables us to obtain an optimal solution for almost-linear instances in $O(2^{\abs{\mathcal{N'}}}\frac{N}{\varepsilon^2})$.

%% file: Sections/Stability.tex
\section{Coalition Stability and Membership Incentives} \label{sec: dynamics}

In this section, we aim to provide further insights on the optimal coalition. We analyze stability (\Cref{subsec:stability}), retention of ``near-zero'' contributes (\Cref{subsec:retention}), and gain insights on the size of optimal coalitions (\Cref{subsec:VSR}). 
\subsection{Stability}\label{subsec:stability}
To understand whether Principal is satisfied with the optimal coalition, we use the concept of stable coalition. 
\begin{definition}[Stable coalition]\label{def: stable coalition}
A Coalition $\mathcal{C}$ is stable if for every $\mathcal{C}' \subset \mathcal{C}$ it holds that
\begin{align*}
W(\sigma^\star(\mathcal{C}), \mathcal{C}) \geq W(\sigma^\star(\mathcal{C}), \mathcal{C}').
\end{align*}
\end{definition}
Another way to think about stable coalition is using the \emph{simultaneous move game}, in which  Principal chooses a coalition while the users choose their effort profile. A coalition is stable if Principal has no beneficial unilateral deviation. If the optimal coalition $\mathcal C^\star$ is stable, Principal cannot benefit from removing a member of $\mathcal C^\star$. We remark that a stable coalition always exists, as the empty coalition is stable by definition. A stable coalition is desirable because, in practice, instability can lead to regret, repeated membership changes, and lower utility. Unfortunately, the next proposition shows that the optimal coalition need not be stable.
\begin{proposition} \label{find coalition not stable}
There are instances where $\mathcal{C}^\star$ is not a stable coalition.
\end{proposition}

\subsection{Retention Incentive}\label{subsec:retention}
This subsection shows that retaining a seemingly low-contribution player can be optimal due to effort complementarities. In particular, a player's presence may sustain higher effort by others; removing that player can trigger lower best responses by the remaining members, reducing the shared benefit and Principal's utility.
\begin{proposition}\label{keep low performing}
For every $\varepsilon > 0$ there exists an instance where $i\in \mathcal{C}^\star$ and $s_i(\sigma^\star_i(\mathcal{C}^\star)) - s_i(0) < \varepsilon$.
\end{proposition}
\Cref{keep low performing} implies that a ``near-zero'' contributor can be pivotal for maintaining high effort in the coalition. %Therefore, while Principal would myopically want to remove that player as they render the coalition unstable, this may cause 

\subsection{Marginal Value‑to‑Share Ratio}\label{subsec:VSR}
To assess whether a player's membership is justified for a coalition, we use the Value‑to‑Share ratio ($VSR$):
\[
VSR(i, \mathcal{C}) = \frac{f(\sigma^\star(\mathcal{C}), \mathcal{C}) - f(\sigma^\star(\mathcal{C}), \mathcal{C} \setminus \{ i \})}{\theta_i}
\]
The $VSR$ compares $i$'s marginal contribution to the shared benefit with $i$'s portion $\theta_i$. High $VSR$ suggests retaining $i$; low $VSR$ indicates that excluding $i$ may be beneficial. The next result shows that the more players Principal exclude from the coalition, the lower each player's marginal effect gets. 
\begin{proposition}\label{prop:VSR}
For any $\mathcal{C}, \mathcal{C}' \subseteq \mathcal{N}$ and $i \in \mathcal{N}$ such that $i \in \mathcal{C'} \subset \mathcal{C}$, it holds that $VSR(i, \mathcal{C}) \geq VSR(i, \mathcal{C}')$.  
\end{proposition} 
The proof relies on \emph{increasing differences} property of supermodular games: as the coalition shrinks, each remaining member's marginal contribution weakly decreases, so their $VSR$ falls. A practical implication is that optimal coalitions tend to be either large or small: in large coalitions many members exhibit high $VSR$; conversely, once removals begin, $VSRs$ drop and can trigger a cascade of exclusions. Indeed, this is aligned with our simulations (see \Cref{sec:simulations}).

%% file: Sections/simulation.tex
\section{Simulations}\label{sec:simulations}

\begin{figure*}[t]
\centering
\begin{subfigure}[b]{0.417\textwidth} 
\includegraphics[width=\linewidth]{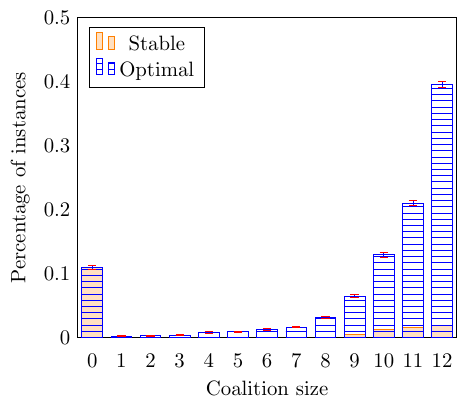}
\caption{} \label{fig sim1}
\end{subfigure}
\hfill
\begin{subfigure}[b]{0.53\textwidth} 
\centering
\includegraphics[width=\linewidth]{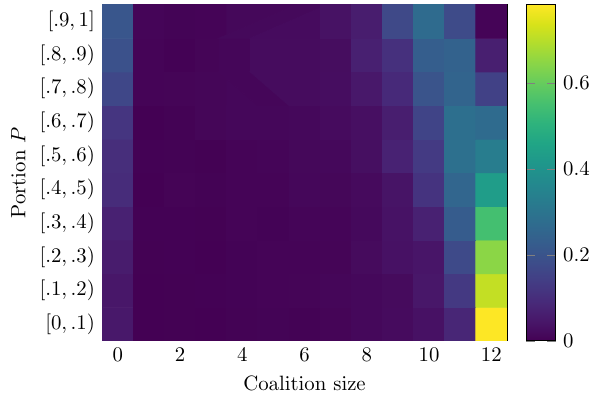}
\caption{} \label{fig sim2}   
\end{subfigure}
\caption{Analysis of coalition size in equilibrium. Figure~\eqref{fig sim1} shows the empirical frequency of optimal coalitions (blue, horizontal lines) versus stable coalitions (orange, solid). 
Empty coalitions and large coalitions (nine to twelve players) occur most frequently. Figure \eqref{fig sim2} provides a more granular view, grouping instances by the total player share $P=\sum_{i=1}^N \theta_i$. 
Each row plots the empirical distribution of optimal coalition sizes within its group. When $P$ is small---meaning the Principal captures nearly the entire surplus, the full coalition frequently emerges in equilibrium. 
As $P$ increases, the empty coalition becomes increasingly prevalent.}
\end{figure*}

In this section, we provide a suite of simulations to understand the structure of optimal coalition (at equilibrium), as well as its dynamics.

\paragraph{Generating instances}
To demonstrate the dynamics of the game, we focus on instances in which the shared benefit exhibits a high degree of coupling between players, implemented through multiplicative interactions among their contribution functions. To that end, we focus on instances with twelve players. Each instance is constructed using a shared benefit function which is a product over all the contributions, namely $F(s_1, \ldots, s_N) = \prod_{i=1}^N s_i$. Each player's contribution function is given by $s_i(e_i, g_i) = \sqrt{\alpha_i e_i + \beta_i g_i}$, and their cost function is linear: $c_i(e_i) = \delta_i e_i$. The coefficients $\alpha_i$, $\beta_i$, and $\delta_i$ are sampled independently and uniformly from the interval $[0, 1]$.

To determine the portion of the shared benefit each player receives, we first sample a value $P \sim U([0, 1])$, representing the total portion allocated to all players collectively. Then, for each player, we sample $t_i$ independently and uniformly from $[0, 1]$, and set the share of player $i$ to be $\theta_i = P \cdot \frac{t_i}{\sum_{j = 1}^N t_j}$. This construction ensures that including all players in the coalition can yield positive utility for Principal.

\paragraph{Computing equilibrium profiles.} 
We employ a brute-force search to identify the optimal coalition for Problem~\eqref{find coalition} (The algorithm of Section~\ref{subsec: alg} is efficient only when the shared-benefit function is linear or nearly linear, which is not the case under the strong coupling of $F$). For each possible coalition, we calculate the corresponding $\domeqn$ equilibrium profile using a best-response dynamics~\cite{topkis1998supermodularity} with a tolerance of $10^{-8}$ (required since the effort space is continuous).

\paragraph{Hardware} We used a standard PC with intel Core i7-9700k CPU and 16 GB RAM for running the simulations. The entire execution took roughly 15 hours.

\subsection{Results}
We report the results over 10,000 generated instances. 

\paragraph{Coalition size}
Figure~\eqref{fig sim1} shows the empirical frequency of coalition sizes. For optimal coalitions (blue with horizontal line pattern), we see that coalitions of medium size (1-8 players) are very rare, while the empty coalition and coalitions with 10-12 players are more common (the latter are chosen $70\%$ of the time). The orange (solid fill) bars represent the proportion of stable optimal coalitions. By definition, the empty coalition is always stable. Surprisingly, while one-player coalitions are rare ($26$ out of $10,000$), they are all stable. Roughly $5\%$ of the optimal full coalitions are stable, with as many as $10\%$ stable optimal coalitions for ten-player coalitions.

Figure~\eqref{fig sim2} provides a more granular view of the optimal coalition size as a function of the total proportions of the players $P=\sum_{i=1}^N \theta_i$ (recall we sample $P\sim U[0,1]$). This helps us analyze how sensitive the results in Figure~\eqref{fig sim1} are to $P$. The horizontal axis of the heatmap is the coalition size, and the vertical axis is the $P$ interval. Each row in the heatmap is the aggregated empirical frequency distribution. 
Lighter colors represent higher probability mass, and dark colors represent lower probability mass. The empty coalition becomes more common as $P$ increases. In contrast, the full coalition is more frequent for low values of $P$. As $P$ increases, medium-sized coalitions are more frequent: the full coalition is too costly for Principal.

\paragraph{Myopic removal dynamics} 
While our equilibrium analysis offers a forward-looking perspective, we also want to take into account stability, e.g., like the \emph{trembling hand} effect~\cite{marchesi2021trembling,farina2018practical}. In our context, even if the optimal coalition and associated player strategies have materialized, Principal can still be tempted to remove ineffective players. What would be a stable outcome?

We conduct additional analysis employing the \emph{myopic removal dynamics}, defined as follows. The initial state is the optimal coalition $\mathcal C^\star$ and its $\domeqn$ equilibrium $\sigma^\star(\mathcal C^\star)$. If the coalition is not stable, Principal deviates to a stable coalition that maximizes their utility (assuming unchanged effort levels). We thus obtain a new coalition $\mathcal C'$, and then players deviate to the associated $\domeqn$ equilibrium $\sigma^\star(\mathcal{C}')$. This process repeats until convergence to a stable coalition.

Roughly $80\%$ of the instances end up with the empty coalition, while converging to coalitions with 1-2 or 8-12 players are probable but less frequent (each accounting for roughly $0.08\%$ of instances). Indeed, this is aligned with our intuition from Subsection~\ref{subsec:VSR} about decreasing $VSR$ values. We report the full results of this analysis in the appendix.

%% file: Sections/discussion.tex
\section{Discussion and Future Work}
This paper studies how the presence of GenAI changes team dynamics in strategic environments.  GenAI impacts strategic players working towards a shared task in several ways. First, while hypothetically players could use GenAI to be more productive or improve the quality of their work, it can also encourage players to stop investing effort, which may be detrimental to the shared benefit. These effects are exacerbated in strategic environments. Moreover, since GenAI is a tool available to everyone, it allows a layperson without domain expertise to replace existing workers, which tremendously affects coalition formation.
From a computational perspective, finding the optimal coalition is challenging.

We identify several interesting directions for future work. First, we have considered exogenously given portions $(\theta_i)_i$. Despite this assumption holding in many scenarios, e.g., granting options to start-up employees following industry standard, other cases exist. Future work could study how to vary the portions to reach a better equilibrium outcome. Another direction is to study contract design, where players must contribute significantly more than their GenAI replacement. Future work could also explore more diverse forms of interaction around the shared benefit, such as competition or substitutable relationships.

%% file: appendix.tex
\section{Proofs Omitted From Section~\ref{sec: free ride}} \label{appn: free ride}

\subsection{Supermodular games}
We now present the formal definition of supermodular games~\cite{topkis1998supermodularity}.

\begin{definition}
Let $(\mathcal{N}, A, U)$ be a game where $\mathcal{N}$ is the set of players, $A = \prod_{i \in \mathbb{N}}A_i$ is the space of feasible joint strategies and $U_i : A_i \rightarrow \mathbb{R}$ is the payoff of player $i$. The game is supermodular if:
\begin{enumerate}
    \item $A$ is a sublattice of $\prod_{i \in \mathbb{N}} \mathbb{R}^{N}$.
    \item $U_i(a_i, \textbf{a}_{-i})$ is supermodular in $a_i$ on $A_i$ for each $\textbf{a}_{-i}$ on $A_{-i}$ and every player $i$.
    \item $U_i(a_i, \textbf{a}_{-i})$ has increasing differences in $(a_i, \textbf{a}_{-i})$ on $A_i \times A_{-i}$.
\end{enumerate}
\end{definition}

\subsection{Proofs Omitted From Subsection~\ref{subsec: solution concept}}

\begin{proofof}{supermodular game}
In every $\eogame$ game, the action space of each player is the effort space $[0, 1]$. Therefore, the joint strategy space is $[0, 1]^N$ which is a continuous and compact subspace of $\mathbb{R}^N$. Furthermore, given an $\eogame$ game, the GenAI usage profile is fixed and therefore the contribution $s_i$ is only a function of the effort $e_i$. In other words, we have a mapping between an effort profile to the shared benefit function $F$.

Let $F_g(\textbf{e})$ be the mapping between the shared benefit and the effort profile, such that
\[
F_g(\textbf{e}) = F(s_1(e_1, g_1), \ldots, s_N(e_N, g_N)).
\]
Since the joint strategy space is continuous and compact, then according to \citet{vives1990nash}, it holds that if $F_g$ satisfies increasing differences in $(e_i, \textbf{e}_{-i})$ on $[0, 1] \times [0, 1]^{N-1}$ then $F_g$ also satisfies the supermodular property.

Notice that any multilinear polynomial function $h(x_1, \ldots x_N)$ satisfies that for every $i,j \in \mathcal{N}$ it holds that
\begin{align} \label{cont inc diff}
\frac{\partial^2 h}{\partial x_i \partial x_j} \geq 0,    
\end{align}
and so does the shared benefit function. The condition in Equation~\eqref{cont inc diff} is the increasing differences condition for continuous and twice differentiable functions.

Fix any arbitrary $\eogame$ game, and we show that $U_i(\textbf{e})$ also satisfies the increasing differences property. For that, we need to prove that for every $\textbf{e} > \textbf{e}'$ it holds that
\begin{align*}
U_i(e_i, \textbf{e}_{-i}) - U_i(e'_i, \textbf{e}_{-i}) \geq U_i(e_i, \textbf{e}_{-i}') - U_i(e'_i, \textbf{e}_{-i}').
\end{align*}

If $i \notin \mathcal{C}$ then $U_i(e_i, \textbf{e}_{-i}) - U_i(e'_i, \textbf{e}_{-i}) = U_i(e_i, \textbf{e}_{-i}') - U_i(e'_i, \textbf{e}_{-i}')$. Otherwise, we can write the utility explicitly
\begin{align*}
\theta_i F_g(e_i, \textbf{e}_{-i}) - c_i(e_i) - \theta_i F_g(e'_i, \textbf{e}_{-i}) + c_i(e'_i) \geq \theta_i F_g(e_i, \textbf{e}_{-i}') - c_i(e_i) - \theta_i F_g(e'_i, \textbf{e}_{-i}') + c_i(e'_i).
\end{align*}

The costs cancel out, and therefore, it is equivalent to the following condition
\begin{align*}
F_g(e_i, \textbf{e}_{-i}) - F_g(e'_i, \textbf{e}_{-i}) \geq F_g(e_i, \textbf{e}_{-i}') - F_g(e'_i, \textbf{e}_{-i}').
\end{align*}
This is the increasing differences condition of $F_c$, which trivially holds since $F_c$ satisfies Equation~\eqref{cont inc diff}. Thus $U_i$ also satisfies the increasing differences. For the same reason, $U_i$ also satisfies the supermodularity property and therefore any $\eogame$ game is a supermodular game. This concludes the proof of \Cref{supermodular game}
\end{proofof}

\begin{proofof}{obs: increasing equilibria}
If player $i \notin \mathcal{C}$ then $U_i$ is only a function of $e_i$. Furthermore, it decreases as a function of $e_i$. Therefore, in every equilibrium profile, it holds that $e_i = 0$ and $U_i(\textbf{e}, \textbf{g}, \mathcal{C}) = U_i(\textbf{e}', \textbf{g}, \mathcal{C})$

If player $i \in \mathcal{C}$ then it holds that
\begin{align*}
U_i(e'_i, \textbf{e}'_{-i}, \textbf{g}, \mathcal{C}) = f(e'_i, \textbf{e}'_{-i}, \textbf{g}, \mathcal{C}) - c_i(e'_i) \leq f(e'_i, \textbf{e}_{-i}, \textbf{g}, \mathcal{C}) - c_i(e'_i) \leq f(e_i, \textbf{e}_{-i}, \textbf{g}, \mathcal{C}) - c_i(e_i) = U_i(e_i, \textbf{e}_{-i}, \textbf{g}, \mathcal{C})
\end{align*}

Where the first inequality follows since $F$ is non-decreasing in $s_j$ and $s_j(e_j, g_j)$ is non-decreasing in $e_j$ for every player $j \in \mathcal{C}$. The second inequality follows since $\textbf{e}$ is an equilibrium and therefore $e_i$ is the best response of player $i$ to the profile $\textbf{e}_{-i}$. This concludes the proof of \Cref{obs: increasing equilibria}.
\end{proofof}

\subsection{Proofs Omitted From Subsection~\ref{subsec: low effort inducing}}

\begin{proofof}{obs: example compare}
We start with the $\eogame$ game induced by $\textbf{g} = \onevec$. The utilities of the players are given by
\begin{align*}
\begin{cases}
& U_1(e_1, e_2, \onevec, \mathcal{N}) = 2.4 \sqrt{e_1} \sqrt{e_2 + 0.2} - \ln(1+e_1) \\
& U_2(e_1, e_2, \onevec, \mathcal{N}) = 2.4 \sqrt{e_1} \sqrt{e_2 + 0.2} - 3\ln(1+e_2)
\end{cases}.
\end{align*}

Player $1$ has a dominant strategy, which is to invest effort $e_1$. That is. for every $e_2 \geq 0$ it holds that $1 = \argmax_{e_1} U_1(e_1, e_2, \onevec, \mathcal{N})$. Therefore, player $2$ solves an optimization problem which depends only on $e_2$, where the best response is $0 = \argmax_{e_2} U_2(1, e_2, \onevec, \mathcal{N})$. Hence, the equilibrium is $\sigma^\star(\onevec, \mathcal{N}) = (e_1, e_2) = (1, 0)$, for which the utilities are $U_1(\sigma^\star(\onevec, \mathcal{N}), \onevec, \mathcal{N}) = 0.38$ and $U_2(\sigma^\star(\onevec, \mathcal{N}), \onevec, \mathcal{N}) = 1.07$.

We now move on to the $\eogame$ game induced by $\textbf{g} = \zerovec$. In this game, the utilities are
\begin{align*}
\begin{cases}
& U_1(e_1, e_2, \onevec, \mathcal{N}) = 2.4 \sqrt{e_1} \sqrt{e_2} - \ln(1+e_1) \\
& U_2(e_1, e_2, \onevec, \mathcal{N}) = 2.4 \sqrt{e_1} \sqrt{e_2} - 3\ln(1+e_2)
\end{cases}.
\end{align*}

Observe that $s_2(0, 1) = s_2(0.2, 1)$ and therefore if $e_2 \geq 0.2$ then player $1$'s best response is $e_1 = 1$. Furthermore, the best response of player $2$ for player $1$ investing effort $e_1=1$ is $0.25 = \argmax_{e_2} U_2(1, e_2, \zerovec, \mathcal{N})$. Thus, the equilibrium is $\sigma^\star(\zerovec, \mathcal{N}) = (1, 0.25)$ for which the utilities are $U_1(\sigma^\star(\zerovec, \mathcal{N}), \zerovec, \mathcal{N}) = 0.5$ and $U_2(\sigma^\star(\zerovec, \mathcal{N}), \zerovec, \mathcal{N}) = 0.53$. This concludes the proof of \Cref{obs: example compare}
\end{proofof}

\begin{proofof}{GenAI induced no effort}
Consider the following $2$-players game: $\theta_1 = \theta_2 = 0.25$, $s_1(e_1, g_1) = e_1$, $s_2(e_2, g_2) = \sqrt{e_2 + \varepsilon g_2}$, $F(s_1, s_2) = 4s_1 s_2$ $c_1(e_1)$ with the costs $c_1(e_1) = 0$ and $c_2(e_2) = \sqrt{e_2}$. We fix coalition $\mathcal{C} = \mathcal{N}$ and look at the $\eogame$ games induced by coalition $\mathcal{C}$ with effort profiles $\onevec$ and $\zerovec$.

The utilities in the $\eogame$ game induced by $\textbf{g} = \zerovec$ are
\begin{align*}
\begin{cases}
& U_1(e_1, e_1, \zerovec, \mathcal{N}) = e_1 \sqrt{e_2} \\
& U_2(e_1, e_1, \zerovec, \mathcal{N}) = e_1 \sqrt{e_2} - \sqrt{e_2} = \sqrt{e_2}(e_1 - 1).
\end{cases}
\end{align*}

In this game, player $1$ is indifferent between $g_1 = 0$ and $g_1 = 1$. Furthermore, player $1$ incurs no cost; therefore, the dominant strategy is to choose $e_1 = 1$. In this case, player $2$ receives no utility for every effort $e_2 \in [0, 1]$. Hence, any profile in $\{ (1, e_2) \mid e_2 \in [0, 1]\}$ is an equilibrium, while $\domeqn$ profile is $\sigma^\star(\zerovec, \mathcal{N}) = (1,1)$.

We move on to the $\eogame$ induced by $\textbf{g} = \onevec$. The utilities in this game are given by
\begin{align*}
\begin{cases}
& U_1(e_1, e_2, \onevec, \mathcal{N}) = e_1 \sqrt{e_2 + \varepsilon} \\
& U_2(e_1, e_2, \onevec, \mathcal{N}) = e_1 \sqrt{e_2 + \varepsilon} - \sqrt{e_2}
\end{cases}.
\end{align*}

Similar to before, player $1$ always chooses $e_1 = 1$. In the case of player $2$, the dynamics now change. Player $2$ is no longer indifferent to its effort level, since the utility is now a decreasing function
\[
\frac{d}{de_2} U_2(1, e_2, \onevec, \mathcal{N}) = \frac{1}{2} \left( \frac{1}{\sqrt{e_2 + \varepsilon}} - \frac{1}{\sqrt{e_2}} \right).
\]
Therefore, player $2$'s optimal effort is $e_2 = 0$. Hence, there is only one equilibrium, which is also the $\domeqn$ profile $\sigma^\star(\onevec, \mathcal{N}) = (1, 0)$.
This concludes the proof of \Cref{GenAI induced no effort}.
\end{proofof}

\subsection{Proofs Omitted From Subsection~\ref{subsec: price of GenAI}}

\begin{proofof}{pog result}
We use the instance of \Cref{GenAI induced no effort} with $\mathcal{C} = \mathcal{N}$. Recall that $F(e_1, e_2, g_1, g_2) = e_1\sqrt{e_2 + 0.2g_2}$, $\sigma^\star(\zerovec, \mathcal{C}) = (1, 1)$ and $\sigma^\star(\onevec, \mathcal{C}) = (1, 0)$, therefore
\begin{align*}
\pog = \frac{f(\sigma^\star(\zerovec, \mathcal{C}), \zerovec, \mathcal{C})}{f(\sigma^\star(\onevec, \mathcal{C}), \onevec, \mathcal{C})} = \frac{1\sqrt{1}}{1\sqrt{\varepsilon}} = \frac{1}{\sqrt{\varepsilon}}.
\end{align*}
This concludes the proof of \Cref{pog result}.
\end{proofof}

\section{Proofs Omitted From Section~\ref{sec: replace}}
\subsection{Proofs Omitted from Subsection~\ref{subsec: two stage hardness}}

\begin{proofof}{find coalition np-hard}
We show that Problem~\eqref{find coalition} is NP-complete using a reduction from the clique problem.
Formally, the clique problem is the following: given an undirected graph $\mathcal{G}=(V, E)$ and a scalar $k$, is there a subset of vertices of at least size $k$ such that an edge connects every two vertices in the subset? We first show our construction, that is, an instance of our problem which represents the clique problem, constructed in polynomial time. Then, we analyze the behavior of each player in the $\domeqn$ equilibrium. We further split this step into two stages: first, given a coalition, we analyze the behavior of every player $i > 1$, and then we analyze the decision of player $1$. Lastly, we show that the clique problem has a clique of size $k$ only if our problem has a coalition of at least $k$ players.

\paragraph{Step 1: Instance construction}
We define a game of $N = \left| V \right|$ players. Each vertex in $V$ represents a player in our game. Let $\theta_i = \frac{2k - 1}{k(3k - 2)}$ for every player $i \in \mathcal{N}$. Furthermore, let $s_i(e) = \mathds{1}_{\{e = 1\}}$ for every player $i \in \mathcal{N}$ while there is no cost for the players $c_i(e) = 0$. Next, we define the shared benefit $F$ according to the edges in $\mathcal{G}$. $F$ has a term of the form $s_i(e_i) s_j(e_j)$ for every edge $(i, j)$. Formally
\[
F(e_1, \ldots e_N) = \sum_{(i, j) \in E} s_i(e_i) s_j(e_j).
\]

\paragraph{Step 2: Analysis of the dominant equilibrium}

The utility of each player is given by
\[
U_i(\textbf{e}, \textbf{g}, \mathcal{C}) = \mathds{1}_{ \{ i \in \mathcal{C} \} } \frac{2k - 1}{k(3k - 2)} \sum_{(l, j) \in E} \mathds{1}_{ \{ l \in \mathcal{C} \} }, \mathds{1}_{ \{ j \in \mathcal{C} \} } s_l(e_l) s_j(e_j) 
\]

Since the players incur no cost, they have a dominant strategy, which is to always choose $e_i = 1$. Therefore, throughout this proof, we analyze the strategy $\textbf{e} = \onevec$.

We now move on to Principal. The utility of Principal is given by
\begin{align*}
W(\onevec, \onevec, \mathcal{C}) &= \left( 1 - \sum_{i \in \mathcal{C}} \theta_i \right) \sum_{(i, j) \in E} \mathds{1}_{ \{ i \in \mathcal{C} \} }, \mathds{1}_{ \{ j \in \mathcal{C} \} } s_i(e_i) s_j(e_j) \\
&= \left( 1 - \left| \mathcal{C} \right| \frac{2k - 1}{k(3k - 2)} \right) \sum_{(i, j) \in E} \mathds{1}_{ \{ i \in \mathcal{C} \} }, \mathds{1}_{ \{ j \in \mathcal{C} \} }.
\end{align*}

Notice that the summation $\sum_{(i, j) \in E} \mathds{1}_{ \{ i \in \mathcal{C} \} }, \mathds{1}_{ \{ j \in \mathcal{C} \} }$ has an element for every edge between the vertices such that their represented players are in the coalition. Therefore, it is upper bounded by the case where all the vertices (which are represented in the coalition) share an edge with one another. This is exactly the definition of a clique of size $\left| \mathcal{C} \right|$, and formally
\[
\sum_{(i, j) \in E} \mathds{1}_{ \{ i \in \mathcal{C} \} }, \mathds{1}_{ \{ j \in \mathcal{C} \} } \leq \binom{\left| \mathcal{C} \right|}{2} = \frac{\left| \mathcal{C} \right|(\left| \mathcal{C} \right| - 1)}{2}.
\]

\paragraph{Step 3: equivalence of solutions}

We define $R(\tilde{k}) = \left( 1 - \tilde{k} \frac{2k - 1}{k(3k - 2)} \right) \frac{\tilde{k}(\tilde{k} - 1)}{2}$ and notice that $W(\onevec, \onevec, \mathcal{C}) = R(\left| \mathcal{C} \right|)$.

We now use the following lemma to show that $R(\tilde{k})$ has a maximum point at $\tilde{k} = k$.
\begin{lemma} \label{clique reduction utility function}
For every $k' \in \mathbb{N}$ it holds that $R(k) > R(k')$.
\end{lemma}

\Cref{clique reduction utility function} implies that the maximum utility Principal can achieve is
\[
W_{max} = R(k) = \left( 1 - k \frac{2k - 1}{k(3k - 2)} \right) \frac{k(k - 1)}{2} = \frac{k(k - 1)^2}{6k-4}
\]

Therefore, we now show that given a graph $\mathcal{g}$, the solution to the clique problem, of deciding whether there is a clique of at least size $k$ is equivalent to asking whether $\max_\mathcal{C} W(\onevec, \onevec, \mathcal{C})$ is equal to $W_{max}$ on the constructed instance.

We first show that there is a clique of size $k$ if and only if $\max_\mathcal{C} W(\onevec, \onevec, \mathcal{C}) = W_{\max}$. If a clique has a solution of at least size $k$, then there exists a sub-clique in it of size $k$. By definition of a clique, every two vertices in it are connected. Therefore, let $\mathcal{V}$ be the clique of size $k$ and it holds that $W(\onevec, \onevec, \mathcal{V}) = R(\left| \mathcal{V} \right|) = R(k) = W_{\max}$.

If there exists a solution such that $W(\onevec, \onevec, \mathcal{C}) = W_{max}$ then since $R(k)$ is the highest value of $R(\tilde{k})$ for $\tilde{k} > 0$ and $\tilde{k} = k$ is the only point which achieves that value it holds that $\left| \mathcal{C} \right| = k$. Furthermore, by construction, $R(\tilde{k})$ represents a clique of size $\tilde{k}$ and therefore $\mathcal{C}$ has to be a clique of size $k$.

Lastly, the opposite is also true. If $\mathcal{C}$ is not a clique of size $k$, then $W(\onevec, \onevec, \mathcal{C}) < W_{\max}$. This concludes the proof of \Cref{find coalition np-hard}.
\end{proofof}

\begin{proofof}{clique reduction utility function}
We begin by taking the derivative.

\begin{align*}
\frac{d}{d\tilde{k}} R(\tilde{k}) &= \frac{d}{d\tilde{k}} \left( \left( \tilde{k}^2 - \tilde{k} \right)\frac{1}{2} - \left(\tilde{k}^3 - \tilde{k}^2 \right) \frac{2k-1}{2k(3k - 2)} \right) \\
&= \tilde{k} - 0.5 - \frac{3}{2}\tilde{k}^2 \frac{2k-1}{k(3k-2)} + \tilde{k} \frac{2k-1}{k(3k-2)}.
\end{align*}

Observe that for $\tilde{k} = k$ it holds that
\begin{align*}
\frac{d}{d\tilde{k}} R(k) &= k - 0.5 - \frac{3}{2}k^2 \frac{2k-1}{k(3k-2)} + k \frac{2k-1}{k(3k-2)} \\
&= k - 0.5 - \frac{3}{2}k \frac{2k-1}{3k-2} + \frac{2k-1}{3k-2} \\
&= k - 0.5 - \frac{3k^2 - 1.5k - 2k + 1}{3k-2} \\
&= \frac{3k^2 - 2k - 1.5k + 1 - 3k^2 + 3.5k - 1}{3k-2} = 0.
\end{align*}
Furthermore, the second derivative at $\tilde{k} = k$ is given by
\begin{align*}
\frac{d^2}{d\tilde{k}^2} R(k) = 1-3k \frac{2k-1}{k(3k-2)} + \frac{2k-1}{k(3k-2)} = \frac{3k^2 - 2k - 6k^2 + 3k + 2k - 1}{k(3k+2)} = \frac{-3k^2 + 3k - 1}{k(3k+2)},
\end{align*}
which is negative and therefore $\tilde{k} = k$ is a local maximum of $R(\tilde{k})$.

Similarly, the point $\tilde{k} = \frac{3k-2}{3(2k-1)}$ is the other extreme point, which is a local minimum. Next, notice that $\frac{3k-2}{3(2k-1)} < 1$ for every $k > 1$, thus $R(\tilde{k})$ is an increasing function in $\tilde{k} \in [1, k]$ and a decreasing function for every $\tilde{k} > k$. Since $\tilde{k} = k$ is a local maximum point, it holds that $R(k) > R(k')$ for every $k' \geq 1$.

Finally, we show for $\tilde{k} = 0$ that
\begin{align*}
R(0) = -\frac{k(k-1)^2}{6k-4} < 0 = R(k)
\end{align*}
This concludes the proof of ~\Cref{clique reduction utility function}.
\end{proofof}

\subsection{Proofs Omitted From Subsection~\ref{subsec: alg}}
Here we present the algorithm for computing the optimal coalition in \Cref{alg for finding coalition}.

\begin{algorithm}[t]
\textbf{Input:} $G$, $\varepsilon$. \\
\textbf{Output:} $\mathcal{C}^\star$
\begin{algorithmic}[1]
\small % add this command to decrease the font size
\caption{Find Coalition Outer Problem (FCOP)} \label{alg: outer}

\STATE Let $\mathcal{W}[k] \gets 0$ for every $k \in \{ 0, 1, \ldots, \left \lceil \frac{1}{\varepsilon} \right \rceil \}$
\STATE Let $\hat{s}_i = \gamma_i \left(s_i^\star - s_i(0)\right)$

\FOR{$k \in \{ 0, 1, \ldots, \left \lceil \frac{1}{\varepsilon} \right \rceil \}$}
    \STATE $\mathcal{C}_k \gets Find\_k\_coalition((\theta_i)_i, (\hat{s}_i)_i, \varepsilon, k)$ \label{line call knapsack}
    \STATE $\mathcal{W}[k] \gets (1-k\varepsilon) \left( \sum_{i \in N} \gamma_i s_i(0) + \sum_{i \in \mathcal{C}_k} \hat{s}_i \right)$
\ENDFOR

\RETURN{$\mathcal{C}^\star = \mathcal{C}_k$ associated with $\max_{k} \mathcal{W}[k] $} \;
\end{algorithmic}
\end{algorithm}

The function $Find\_k\_coalition$ is the dynamic programming solution of knapsack for the problem of budget $k\varepsilon$, and $N$ such that each item has weight $\theta_i$ and value $\hat{s}_i$.

We now move on to prove \Cref{alg for finding coalition}.

\begin{proofof}{alg for finding coalition}
To prove this result, we have two requirements to show:
\begin{enumerate}
    \item The optimal solution of the Problem~\eqref{find budget k} satisfies the constraint of the Problem~\eqref{eq:p2 inner} with strict equality.
    \item Algorithm~\ref{alg: outer} returns the optimal solution. Afterward to show the running time.
\end{enumerate}

We start with the first requirement.
Assume in contradiction that there exists a solution to Problem~\eqref{find budget k} $\mathcal{C}^\star$ and $k^\star$ such that $\sum_{i \in \mathcal{C}^\star} \theta_i < k^\star \varepsilon$. Since $\theta_i \in \{ 0, \varepsilon, \ldots, 1\}$ for every $i \in \mathcal{N}$ then there exists $\alpha_i \in \mathbb{N}$ such that $\theta_i = \alpha_i \varepsilon$. Therefore, it there exists $\beta \in \mathbb{N}$ such that 
\[
\sum_{i \in \mathcal{C}^\star} \theta_i = \sum_{i \in \mathcal{C}^\star} \alpha_i \varepsilon = \varepsilon \sum_{i \in \mathcal{C}^\star} \alpha_i = \varepsilon \beta.
\]

Furthermore, from our assumption it holds that $\varepsilon \beta < \varepsilon k^\star$. Hence $(1-k\varepsilon) f(\sigma^\star(\onevec, \mathcal{C}^\star), \onevec, \mathcal{C}^\star) < (1-\beta\varepsilon) f(\sigma^\star(\onevec, \mathcal{C}^\star), \onevec, \mathcal{C}^\star)$. In other words, $k^\star$ is not optimal. 
In addition, there can be no other $\mathcal{C}'$ such that $\sum_{i \in \mathcal{C}'} \theta_i = \beta \varepsilon < k\varepsilon$ and results in higher shared benefit then $\mathcal{C}^\star$. Thus, if $\mathcal{C}^\star$ is optimal, then it must be relative to $k = \beta$, which means that the constraint is satisfies with a strict equality.

We now move on to the correctness of our algorithm. 
First, we use the following lemma
\begin{observation} \label{knapsack correctness}
for every $k \in \{0, 1, \left \lceil \frac{1}{\varepsilon} \right \rceil \}$ it holds that $\mathcal{W}[k]$ in Line~\ref{line call knapsack} is the optimal solution of the Problem~\eqref{eq:p2 inner}.
\end{observation}
The correctness and the proof of \Cref{knapsack correctness} follows directly from the correctness of the dynamic programming solution to the $0-1$ knapsack problem and hence omitted.

Next, from our first requirement we have that the optimal solution of Problem~\eqref{find budget k} is in $\mathcal{W}$. Since there are only finite $\left \lceil \frac{1}{\varepsilon} \right \rceil + 1$ different values that $k$ can get, iterating over all of them is guaranteed to return the optimal solution of Problem~\eqref{find budget k}.

Lastly, we compute the running time of the algorithm. Notice that the function in Line~\ref{line call knapsack} is the dynamic programming solution of the $0-1$ knapsack problem and therefore it runs in $O(\frac{N}{\varepsilon})$. Next, in Algorithm~\ref{alg: outer}, we iterate over $O(\frac{1}{\varepsilon})$ different values for $k$, and in each iteration we call the dynamic programming solution of knapsack. Therefore, the total runtime of the algorithm is $O(\frac{N}{\varepsilon^2})$.
This completes the proof of \Cref{alg for finding coalition}.
\end{proofof}

\section{Proofs Omitted From Section~\ref{sec: dynamics}}

\subsection{Proofs Omitted From Subsection~\ref{subsec:stability}}

\begin{proofof}{find coalition not stable}
Consider the following $2$-players game: Let the portions be $\theta_1 = \theta_2 = 0.4$ and the shared benefit be $F(s_1, s_2) = 2.5s_1 s_2$, where the contribution of each player are given by $s_1(e_1,g_1) = 2e_1$ and $s_2(e_2, g_2) = \sqrt{e_2 + 0.2g_2}$. Let the costs be $c_1(e_1) = \sqrt{e_1}$ and $c_2(e_2) = 1.8e_2$. Therefore, the utilities in this game are given by
\begin{align*}
U_1(e_1,e_2, g_1, g_2, \mathcal{C}) = \begin{cases}
   2e_1 \sqrt{e_2 + 0.2g_2} - \sqrt{e_1} & \mbox{$\mathcal{C} = \mathcal{N}$} \\
   2e_1 \sqrt{0.2} - \sqrt{e_1} & \mbox{$\mathcal{C} = \{ 1 \}$} \\
   - \sqrt{e_1} & \mbox{$1 \notin \mathcal{C}$}
\end{cases},
\end{align*}

\begin{align*}
U_2(e_1, e_2, g_1, g_2, \mathcal{C}) = \begin{cases}
    2e_1 \sqrt{e_2 + 0.2g_2} - 1.8 e_2 & \mbox{$\mathcal{C} = \mathcal{N}$} \\
    - 1.8 e_2 & \mbox{Otherwise}
\end{cases},
\end{align*}

\begin{align*}
W(e_1, e_2, g_1, g_2, \mathcal{C}) = \begin{cases}
    0.5 \cdot 2e_1 \sqrt{e_2 + 0.2g_2} & \mbox{$\mathcal{C} = \mathcal{N}$} \\
    1.5 \cdot 2e_1 \sqrt{0.2} & \mbox{$\mathcal{C} = \{ 1 \}$} \\
    0 & \mbox{$1 \notin \mathcal{C}$}
\end{cases}.
\end{align*}
For each coalition, we calculate the utilities of the players and Principal. For the coalition $\mathcal{C} = \mathcal{N}$, we use the following lemma.

\begin{lemma} \label{lemma domeq full coalition}
The $\domeqn$ equilibrium of the $\eogame$ game induced by $\mathcal{C} = \mathcal{N}$ and $\textbf{g} = \onevec$ is $\sigma^\star(\onevec, \mathcal{N}) = (1, 0.108)$.
\end{lemma}

Using \Cref{lemma domeq full coalition}, we can calculate Principal's utility
\begin{align*}
W(\sigma^\star(\onevec, \mathcal{N}), \onevec, \mathcal{N}) = 0.5 \cdot 2 \sqrt{0.108 + 0.2} = 0.555
\end{align*}

We move on to the next coalition. For $\mathcal{C} = \{ 1 \}$ we solve an optimization problem, for which the optimal decision of player $1$ is to invest $e_1 = 0$. Player $2$ can only gain negative utility and therefore $\sigma^\star(\onevec, \{ 1 \}) = (0, 0)$. Principal's utility in this case is given by
\begin{align*}
W(\sigma^\star(\onevec, \{ 1 \}), \onevec, \{ 1\}) = 0.
\end{align*}
Notice that for any other coalition, the optimal choice of player $1$ and player $2$ is to invest no effort, and therefore Principal has no utility for those coalitions. In other words, the optimal coalition according to Problem~\eqref{find coalition} is $\mathcal{C}^\star = \mathcal{N}$.

Observe that
\begin{align*}
W(\sigma^\star(\onevec, \mathcal{N}), \onevec, \{ 1\}) = 1.5 \cdot 2 \sqrt{0.2} = 1.341.
\end{align*}
In other words, according to the $\gamename$ game, Principal has the incentive to deviate. But as we saw, deviating would lead to a worse coalition and a worse equilibrium.
This concludes the proof of \Cref{keep low performing}
\end{proofof}

\begin{proofof}{lemma domeq full coalition}
To prove this lemma, we need to show that there is no player who has the incentive to deviate.

We start with player $1$. Given the action of player $2$,  $(e_2, g_2) = (0.108, 1)$, player solves the following optimization problem:

\begin{align*}
e_1^\star = \max_{e_1} 2e_1 \sqrt{0.308} - \sqrt{e_1}.
\end{align*}

We solve this optimization by taking the derivative.
\begin{align*}
\frac{dU_1(e_1, 1, 0.108, 1, \mathcal{N})}{de_1} = 2\sqrt{0.308}- \frac{1}{2\sqrt{e_1}}
\end{align*}

The derivative equals $0$ for $e_1 = \frac{1}{16 \cdot 0.308} \approx 0.2029$. That is, the derivative is negative for every $e_1 < \frac{1}{16 \cdot 0.308}$ and is positive for every $e_1 > \frac{1}{16 \cdot 0.308}$. Therefore, the maximum utility can be attained only at $e_1 = 0$ or $e_1 = 1$. If $e_1 = 0$ then player $1$ gets no utility, and if $e_1 = 1$ then
\begin{align*}
U_1(1, 1, 0.108, 1, \mathcal{N}) = 2\sqrt{0.308} - 1 > 0.
\end{align*}

We now show that player $2$ has no incentive to deviate. Player $g_2 = 1$ is a dominant strategy and therefore we only need to solve the optimization problem for $e_2$. That is,
\begin{align*}
g_2^\star = \max_{e_2} \{2\sqrt{e_2 + 0.2} - 1.8e_2\}.
\end{align*}
This optimization problem can be solved by taking the derivative and setting it to $0$:
\begin{align*}
\frac{1}{\sqrt{e_2 + 0.2}} - 1.8 = 0
\end{align*}
Solving this equation results in $e_2 = \frac{1}{1.8^2} - 0.2 \approx 0.108$ which means that player $2$ has no incentive to deviate.
This concludes the proof of \Cref{lemma domeq full coalition}
\end{proofof}

\subsection{Proofs Omitted From Subsection~\ref{subsec:retention}}

\begin{proofof}{keep low performing}
Let $\varepsilon' = \min \{ \varepsilon, 0.9 \}$. We show that \Cref{keep low performing} holds for $\varepsilon'$. Since $\varepsilon' \leq \varepsilon$, it means that \Cref{keep low performing} holds for $\varepsilon$ as well.

Consider the following instance: Let $N=2$, the contributions be $s_1(e_1, g_1) = e_1$, $s_2(e_2, g_2) = g_2\left( 1-\frac{\varepsilon'}{2} \right) + e_2$ and the costs be $c_1(e_1) = e_1$, $c_2(e_2) = \frac{e_2^2}{2 \varepsilon'}$. Let the portions be $\theta_1 = \theta_2 = 0.25$ and the shared benefit $F(s_1, s_2) = 4s_1 s_2$. Without loss of generality, we consider only $\textbf{g} = \onevec$. Therefore, the utilities are

\begin{align*}
U_1(e_1, e_2, \mathcal{C}) = \begin{cases}
    e_1 (e_2 - \frac{\varepsilon'}{2}) & \mbox{$\mathcal{C} = \mathcal{N}$} \\
    -e_1 \frac{\varepsilon'}{2} & \mbox{$\mathcal{C} = \{ 1 \}$} \\
    -e_1 & \mbox{$1 \notin \mathcal{C}$}
\end{cases},
\end{align*}

\begin{align*}
U_2(e_1, e_2, \mathcal{C}) = \begin{cases}
    e_1 \left(1 - \frac{\varepsilon'}{2} + e_2 \right) - \frac{e_2^2}{2\varepsilon'} & \mbox{$\mathcal{C} = \mathcal{N}$} \\
    - \frac{e_2^2}{2\varepsilon'} & \mbox{Otherwise}
\end{cases},
\end{align*}

\begin{align*}
W(e_1, e_2, \mathcal{C}) = \begin{cases}
    2 e_1 \left( 1 - \frac{\varepsilon'}{2} + e_2 \right) & \mbox{$\mathcal{C} = \mathcal{N}$} \\
    2 e_1 \left( 1 - \frac{\varepsilon'}{2} \right) & \mbox{$\mathcal{C} = \{ 1 \}$} \\
    0 & \mbox{Otherwise}
\end{cases}.
\end{align*}

We calculate the $\domeqn$ equilibrium for each coalition and compare Principal's utility.
\begin{itemize}
    \item $\mathcal{C} = \mathcal{N}$: We show that $(1, \varepsilon')$ is the $\domeqn$ equilibrium.
    First, if Player 2 invests $e_2 > \varepsilon'$ then Player 1's utility becomes $\left(1 + \frac{\varepsilon'}{2} \right)e_1 > e_1$ which is maximized for $e_1 = 1$.
    Next, if Player 1 invests $e_1 = 1$ then Player 2 optimization problem becomes
    \[
    e_2^\star = \argmax_{e_2} 1 - \frac{\varepsilon'}{2} + e_2 - \frac{e_2^2}{2\varepsilon'} = \argmax_{e_2} e_2 - \frac{e_2^2}{2\varepsilon'}.
    \]
    This function has only one extreme point, which can be found by taking the derivative.
    \[
    0 = \frac{d}{de_2}\left( e_2 - \frac{e_2^2}{2\varepsilon'} \right) = 1 - \frac{e_2}{\varepsilon'}.
    \]
    Therefore, the optimal effort of Player 2 is $e_2 = \varepsilon'$. Since Player 1 cannot invest more than $e_1 = 1$, it means that $(1, \varepsilon')$ is the $\domeqn$ equilibrium. Under this profile, Principal's utility is $W(1, \varepsilon_1, \mathcal{N}) = 2\left( 1 + \frac{\varepsilon'}{2} \right) = 1 + \varepsilon'$.
    
    \item $\mathcal{C} = \{ 1 \}$: Each player's utility is decreasing in that player's effort and therefore $e_1 = e_2 = 0$. Therefore, Principal's utility is $W(0, 0, \{ 1 \}) = 0$.

    \item $\mathcal{C} = \{ 2 \}$: Same as before. Thus, $e_1 = e_2 = 0$ and $W(0, 0, \{ 2 \}) = 0$.
    \item $\mathcal{C} = \emptyset$: By definition $W(e_1, e_2, \emptyset) = 0$.
\end{itemize}

There, from this analysis, it holds that Principal's optimal coalition is $\mathcal{C}^\star = \mathcal{N}$. Furthermore, notice that
\[
s_2(\sigma^\star(\mathcal{C}^\star)) - s_2(0) = s_2(\varepsilon') - s_2(0) = \left( 1-\frac{\varepsilon'}{2} \right) + \varepsilon' - \left( 1-\frac{\varepsilon'}{2} \right) = \varepsilon' \leq \varepsilon.
\]
Hence, we showed that $2 \in \mathcal{C}^\star$ and Player 2's marginal contribution compared to GenAI is less than $\varepsilon$. This concludes the proof of \Cref{subsec:retention}.
\end{proofof}

\subsection{Proofs Omitted From Subsection~\ref{subsec:VSR}}

\begin{proofof}{prop:VSR}
We denote $\hat{s}(\mathcal{C}) = (\tilde{s}_j(\sigma^\star(\mathcal{C})))_j$ and denote $\hat{s}_{-i}(\mathcal{C})$ all the entries of $\hat{s}(\mathcal{C})$ except for $i$. Therefore, we need to show that
\begin{align*}
\frac{F(\hat{s}_{-i}(\mathcal{C}), \hat{s}_i(\mathcal{C})) - F(\hat{s}_{-i}(\mathcal{C}), s_i(0))}{\theta_i} \geq \frac{F(\hat{s}_{-i}(\mathcal{C}'), \hat{s}_i(\mathcal{C}')) - F(\hat{s}_{-i}(\mathcal{C}'), s_i(0))}{\theta_i}.
\end{align*}
Since $\theta_i > 0$, we can equivalently show that
\begin{align} \label{eq to show}
F(\hat{s}_{-i}(\mathcal{C}), \hat{s}_i(\mathcal{C})) - F(\hat{s}_{-i}(\mathcal{C}), s_i(0)) \geq F(\hat{s}_{-i}(\mathcal{C}'), \hat{s}_i(\mathcal{C}')) - F(\hat{s}_{-i}(\mathcal{C}'), s_i(0))
\end{align}

By parameterizing the contributions corresponding to the players in $\mathcal{C} \setminus \mathcal{C'}$ of the game $\eogame$ game $\tilde{G} = (\mathcal{G}, \onevec, \mathcal{C}')$, we get a parametrized supermodular game. Therefore, from the non-decreasing property of $F$, the $\domeqn$ equilibrium of the players in $\mathcal{C}$ is non-decreasing in the parameters. In other words, it holds that $\hat{s}(\mathcal{C}) \geq \hat{s}(\mathcal{C}')$.
Furthermore, from the non-decreasing property of $F$ again, it holds that
\begin{align*}
    F(\hat{s}_{-i}(\mathcal{C}'), s_i(\mathcal{C})) \geq F(\hat{s}_{-i}(\mathcal{C}'), s_i(\mathcal{C}'));
\end{align*}
hence, Equation~\eqref{eq to show} is satisfies if
\begin{align} \label{eq increasing diff}
F(\hat{s}_{-i}(\mathcal{C}), \hat{s}_i(\mathcal{C})) - F(\hat{s}_{-i}(\mathcal{C}), s_i(0)) \geq F(\hat{s}_{-i}(\mathcal{C}'), \hat{s}_i(\mathcal{C})) - F(\hat{s}_{-i}(\mathcal{C}'), s_i(0)).
\end{align}
The inequality in Equation~\eqref{eq increasing diff} holds from the increasing differences of $F$. This completes the proof of \Cref{prop:VSR}.
\end{proofof}

\section{Additional Simulations}
\begin{figure*}[t]
\centering
\includegraphics{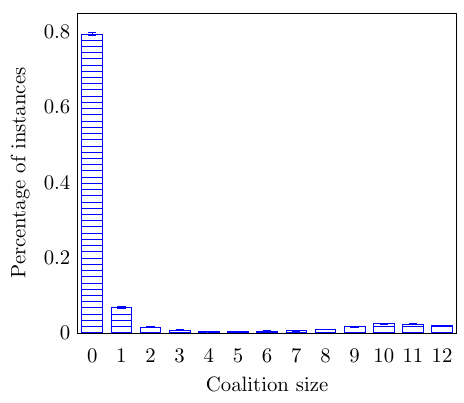}
\caption{Analysis of coalition size after myopic removal due to unstable coalition} \label{dynamic bar graph}
\end{figure*}

In this section, we complement Section~\ref{sec:simulations} with additional simulations. We focus on the myopic removal dynamics. In each instance, we begin with Principal's optimal coalition from Problem~\eqref{find coalition} (as shown in Figure~\eqref{fig sim1}) and let the dynamics evolve. In Figure~\eqref{dynamic bar graph}, we show the empirical frequency of the coalition size that the removal dynamics converged to. While coalitions of ten or more players were optimal in more than $70\%$ of the instances in Figure~\eqref{fig sim1}, less than $7\%$ of those were stable. Furthermore, the result that approximately $80\%$ of the instances ended up with empty coalitions suggests that the $VSR$ ratio of each player is deeply connected to coalition stability. Principal's deviation to a smaller coalition triggers a cascade: the players adjust their efforts downward, which lowers their $VSR$ ratio and further destabilizes the smaller coalition.

